%% file: New_Tools_for_Decomposition.tex
\documentclass[	
		preprint,
		authoryear,
		12pt,
		]{elsarticle}
\usepackage[english]{babel}
\usepackage[latin1]{inputenc}
\usepackage{amsmath, amsfonts, amssymb}
\usepackage{subfigure} 						
\usepackage[usenames,dvipsnames]{xcolor} 			

\usepackage{booktabs} 						
\usepackage{multirow} 						
\usepackage{rotating} 						

\definecolor{blue1}{rgb}{0.0000, 0.5765, 0.8118}

\usepackage{tikz}
\usetikzlibrary{shapes,arrows}

\usepackage[breaklinks,	colorlinks=true]{hyperref} 		
\usepackage[all]{hypcap} 					
\usepackage{wrapfig} 

\usepackage{bbding} 
\usepackage{breakurl}

\newcommand{\good}{{\color{OliveGreen}{\small\CheckmarkBold}}}
\newcommand{\medium}{{\color{Dandelion}{\large$\bullet$}}}
\newcommand{\bad}{{\color{Maroon}{\small\XSolidBrush}}}

\journal{Computers and Geosciences}

\begin{document}

\begin{frontmatter}
	\title{New Tools for Decomposition of Sea Floor Pressure Data\\ 
		\begin{small}A Practical Comparison of Modern and Classical Approaches\end{small}}

	\author[label1]{M. Ehrhardt}
	\address[label1]{Center for Industrial Mathematics, Fachbereich 3, University of Bremen, \\PO Box 33 04 40, D-28334 Bremen, Germany}
	\ead{ehrhardt@math.uni-bremen.de}
	
	\author[label2]{H. Villinger}
	\address[label2]{Department of Geosciences, Fachbereich 5, University of Bremen, \\PO Box 33 04 40, D-28334 Bremen, Germany}
	\ead{vill@uni-bremen.de}
 	
	\author[label1]{S. Schiffler}
	\ead{schiffi@math.uni-bremen.de}

	\begin{abstract}
		In recent years more and more long-term broadband data sets are collected in geosciences. Therefore there is an urgent need of	algorithms which semi-automatically analyse and decompose these data into separate periods which are associated with different processes. Often Fourier and Wavelet Transform is used to decompose the data into short and long period effects but these fail often because of their simplicity. In this paper we investigate the novel approaches Empircial Mode Decomposition and Sparse Decomposition for long-term sea floor pressure data analysis and compare them with the classical ones. Our results indicate that none of the methods fulfils all the requirements but Sparse Decomposition performed best except for computing efficiency. 
	\end{abstract}

\end{frontmatter}
%
%
\input{introduction} 		
\input{methods} 		
\input{results} 		
\input{conclusion}      	
\input{acknowledgements}      	
\input{references} 		
\end{document}

%% file: introduction.tex
\section{Introduction}%
\subsection{Motivation}%
	
Long-term monitoring of earth processes becomes more and more essential and indispensable as only long records will reveal small changes which may eventually have large societal impact in the near future. These processes encompasses geohazards like monitoring of potential landslides and volcanic activity but also processes related to earthquake precursors and climatic signals. In addition real-time observations of tectonic processes with high-precision GPS measurements and satellite based observations of the magnetic and gravity field of the earth give new insights into geological processes at local, regional and plate-size scales. As these signals are mostly small in amplitude and often buried in large-amplitude signals caused by other sources or only visible after major corrections to the field observations have been made it is a major challenge to separate the wanted from the unwanted signal components.%

Broadband data acquisition is now very common as data storage capacity is no major obstacle today and low power data acquisition with autonomous monitoring systems over considerably long time periods (i.e. years) in remote places like the deep ocean floor become more and more feasible. In addition large observational networks - either regionally only temporary like EarthScope (\burl{http://www.earthscope.org/}) - or permanent like Neptune Canada (\burl{http://www.neptunecanada.ca/}) or national or international geodetic GPS observation networks deliver continuously time series whose information content is not completely taken advantage of. All time series contain signal periods from short (several seconds) to long (several month to years); origin and cause of some signal components is sometimes quite well known (deterministic) as e.g. tides from various origins but in a lot of cases especially long-term changes are not well understood. In addition instrumental behaviour like drift and effects of environmental changes severely alter the signal.%

In earth science seismology has faced this problem for a long time which became even more pressing over the last decade during which the number of seismological stations increased accompanied by a tremendous increase of digital storage capacity and a rapid decrease in storage cost. In seismology robust methods were established to extract earthquakes from long data sets automatically and also detect phases and locate earthquakes. This highly automated processing can be achieved because the characteristics of the signals are well-known and years of research have been spend until these algorithms were developed \citep{Bormann2002}.%

Traditional tools for data analysis are mostly based on spectral analysis methods which are well established and available with a lot of software tools. However they tend to introduce unwanted effects especially at short periods due to filters and finite length of time series. Also their requirement of a stationary time series is often not fulfilled in the case of long-term records. Therefore new data exploration tools are needed which enable us to separate the different signal components. These tools should be capable of analysing long time series efficiently and help to identify different signal contributions originating from different processes.%

In this paper we present the results of time series analysis using four different methods. The times series analysed are sea floor pressure records from two different locations (details see section \ref{SECTION:DATA}). The motivation for this investigation came from the fact that the sea floor pressure signal is dominated by a large deterministic tidal pressure signal but due to the high pressure resolution and a technically possible sampling interval of down to 2 seconds, small and short transient signals can be observed. Superimposed on these tides are also long-period transient changes, associated with tectonic and/or oceanographic phenomena but also short period signals originating from local or teleseismic events. The problem was and still is to separate the deterministic from the small-amplitude non-deterministic signals. Traditional tidal analysis methods as described e.g. in \citet{Pawlowicz2002} and available as Matlab\textsuperscript{\textregistered} toolbox (\burl{http://www.eos.ubc.ca/~rich/}) do not always fulfil the requirements.%

We tested four methods in total: the traditional (1) Harmonic Decomposition, (2) Wavelet Decomposition, (3) Empirical Mode Decomposition (EMD) and a novel approach (4) Sparse Decomposition. The last two methods used are rather new and their potential use in time series analysis needs still to be explored. All times series were analysed in the same way with the four methods. After a short description of the methods we present the results and compare the results with respect to their capability of signal decomposition.%

\subsection{Data} \label{SECTION:DATA}%
We used four data sets to test the four methods for time series decomposition. The first data set (SYN) is a synthetic data set which is composed of a tidal signal calculated with a published Matlab\textsuperscript{\textregistered}-script \citep{Pawlowicz2002}, a long-term ramp function and two short-period events. In addition noise with an RMS of 49.16 $Pa$ is added. A step in the middle of the time series simulates a tectonic or oceanographic event. The second time series (MAR) consists of sea floor pressure observations from the Logatchev Hydrothermal Field, located at the Mid-Atlantic Ridge \citep{Gennerich2011}. Sea floor pressure measurements at this site were made in order to monitor the magmatic and hydrothermal activity of the field which might express themselves as changes in sea floor pressure either due to subsidence or uplift or by tremor-like signals related to hydrothermal processes in the subsurface. The third and fourth records (CORK1 and CORK2) come from a so called CORK \citep{Becker2005} installed at the sea floor off Vancouver Island. A CORK (Circulation Obviation Retrofit Kit) is a sea floor installation which is placed on top of a drill hole and isolates the hydrological - and pressure - regime inside the drill hole from the ocean. The main purpose of this installation is the long-term monitoring of pressure signals inside the drill hole in order to identify either slow earthquakes i.e. long period changes in the sub sea floor stress field or the response of the formation to seismic events which reveal the hydrological and permeability structure of the oceanic crust (e.g. \cite{Davis2011}). At every CORK installation sea floor pressures are recorded in addition to the downhole pressures to provide a local pressure reference signal.%

In all cases (MAR, CORK1 and CORK2) pressures were recorded with high-resolution absolute pressure sensors (\burl{http://www.paroscientific.com/uwapp.htm}) with a resolution of 7 $Pa$ which is about 7 $mm$  equivalent water hight change. More details on location and sample intervals can be found in table \ref{tab:data_set_description}.%

\begin{table}[!h]
    \centering
    \caption[Details of the data]{Details of the data sets used in the tests.}
    \begin{tabular}{ccccccc}
	\toprule
	Data set 	& Start 	& End 		& Duration 	& $\Delta t$ 	& \multicolumn{2}{c}{Position}   		\\ 
			&		& 		& $[days]$ 	& $[min]$ 	& Latitude		& Longitude 		\\ \midrule \\
	SYN 		& 		& 		& 41.7	 	& 60 		& $48^\circ$ 26'N	& $128^\circ$ 43'W	\\  
	MAR 		& 01.05.2008 	& 31.05.2008 	& 30 		& 2 		& $14^\circ$ 45'N 	& $44^\circ$ 5' W 	\\ 
	CORK1 		& 26.06.2003 	& 15.09.2005	& 1046		& 10		& \multirow{2}{*}{$48^\circ$ 26'N} & 
												\multirow{2}{*}{$128^\circ$ 43'W}	\\ 
	CORK2 		& 11.09.1996 	& 15.09.2005	& 3292		& 60		& &						\\
	\bottomrule
	\end{tabular} 
	\label{tab:data_set_description}
\end{table}

%% file: methods.tex
\section{Methods for Decomposing}%
In this section we discuss the methods used to decompose the data. These are%
\begin{enumerate}
	\item Harmonic Decomposition
	\item Wavelet Decomposition
	\item Empirical Mode Decomposition
	\item Sparse Decomposition
\end{enumerate}
The classical methods Harmonic and Wavelet Decomposition are well known. Therefore we have chosen them to have a comparison for the new methods Empirical Mode Decomposition and Sparse Decomposition.%

All methods have in common that they seek for a new reprensentation, as a superposition of several \textit{patterns}, of the data, in most cases by basis transformation.

\subsection{Harmonic Decomposition}
	The Harmonic Decomposition uses the Fourier Transform with its efficient implementation as Fast Fourier Transform (FFT). It is known since almost fifty years and used in many data processing routines. The basic idea is to transform the given time signal into the frequency space. In other words we are looking for a reprensentation with sinusiod pattern with different frequencies. We refer to \cite{Groechenig2001} as a reference for a detailed introduction to Fourier Analysis. A much shorter introduction of the Fourier Transform is given in nearly every book about signal analysis, like \cite{Stark2005}.

	Decomposition of our time series with the help of Fourier Analysis starts with a FFT. As we know the frequencies and amplitudes of the tides from theoretical considerations and from observations, we can decompose the data into several components and remove the unwanted and deterministic ones. The remaining spectrum is transformed into a time series which contains only the non-tidal components.

\subsection{Wavelet Decomposition}
	\def\picwidth{0.20\textwidth}
	\def\date{2011-05-13}
	\begin{wrapfigure}[]{r}{4.0cm}
		\centering
		\includegraphics[width=\picwidth]{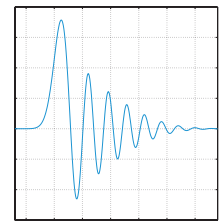} 
		\caption{Wavelet $db40$ (scaling function)}
		\label{explain:db40}
	\end{wrapfigure}
	Another technique for decomposing the data is the Wavelet Decomposition which uses the Wavelet Transform. It is not as old as the Fourier Transform but it is also known very well. Like the Fourier Transform, it transfers the given time signal in another space. This time the basis vectors are scalings and translational displacements of a basic pattern, called \textit{mother wavelet}, hence in this space the coefficients encode when an effect is measured and how long this effect was. Therefore we can again decompose our data in features of a short and long period of time. Just like for the Fourier Transform there are fast implementations for the Wavelet Transform. For further informations about wavelets see \cite{Louis1997} or \cite{Stark2005}.
	
	First of all we have to choose the mother wavelet and a number of levels in which we want to decompose our signal. Unlike the Fourier Transform where the orthogonal functions are pre-defined, we have to choose the mother wavelet a priori. If we have extra knowledge about the data, we can improve the results by choosing a wavelet which is 'similar' to the data or rather the features we would like to extract. In our case the Wavelets \textit{Daubechies 10} - \textit{Daubechies 40} (often abbreviated by $db10$ and $db40$) fit the tidal components very well. After performing the Wavelet Transform with these wavelets, we obtained the best results by using $db40$, which is shown in figure \ref{explain:db40}.

\subsection{Empirical Mode Decomposition}
	The third method we would like to introduce is the Empirical Mode Decomposition and its enhancement the Ensemble Empirical Mode Decomposition (EEMD). The EMD is a new method, invented by \cite{Huang1998}, for decomposing a time signal. Since it is not as basic as the Fourier and Wavelet Transform, we present the algorithm here. The idea is to decompose the given time signal into a finite number of Intrinsic Mode Functions (IMF), which are functions which satisfy two conditions:
	\begin{enumerate}
		\item The number of extrema and the number of zero crossings must differ at most by one.
		\item The mean value of the upper and lower envelope is zero at any time, where the upper envelope is computed by interpolating the maxima and the lower envelope by interpolating the minima.
	\end{enumerate}
	
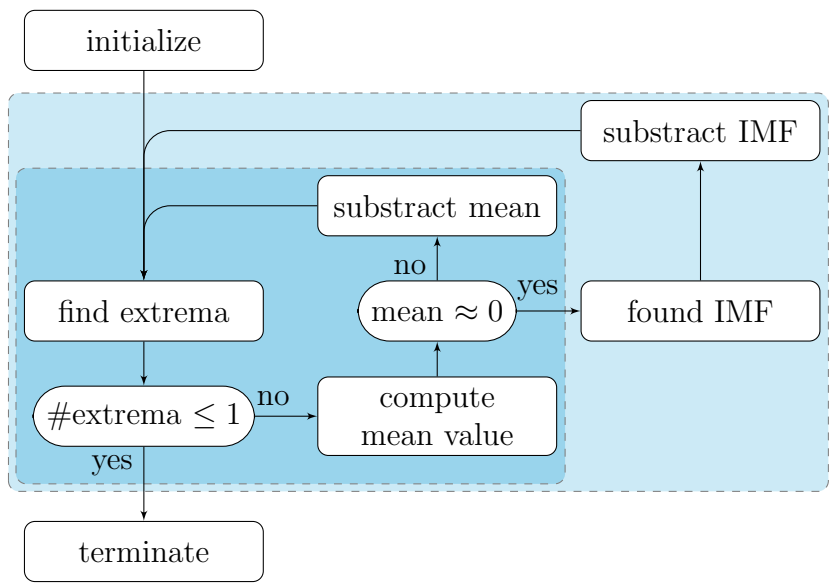
\begin{figure}[ht]
		\centering
		\input{emd_chart}
		\caption{The algorithm of the EMD. The coloured areas represent the loops of the algorithm.}
		\label{explain:EMD:algorithm}
	\end{figure}

	The task of the EMD is to find these IMF's and is done by the following \textit{sifting process}, also shown in figure \ref{explain:EMD:algorithm}.
 	\begin{enumerate}
 		\item Find all extrema of the given data.
 		\item If the number of extrema is one or less, we have found all IMFs and terminate the algorithm.
 		\item Compute the upper and lower envelope by interpolation the maxima and minima and the mean of these.
 		\item If the mean value is zero, we have found an IMF. Start the sifting process again with the data substracted by this IMF. Otherwise start the sifting process again with the data substracted by the mean.
 	\end{enumerate}

	Also the EMD can be seen as a basis transformation, but this time the patterns, which are the IMFs, depend on the data.

	As we have seen we have done no a priori choices but choosing the interpolating scheme, which is predefined by cubic splines. Since abstracting an IMF reduces the number of extrema this algorithm terminates for every finite time signal. A disadvantage of this method is that it is empirical and has no solid theoretical foundation. Compared to the Fourier and Wavelet Transform the computing effort to decompose the data by EMD is higher but still tolerable. A detailed introduction to EMD is given in \cite{Huang1998}.

	A challenge in decomposing time series is to prevent \textit{mode mixing}. Phenomena of similar time scale should be in the same mode (here: IMF) and vice versa phenomena with a totally different scale are expected to have a different mode. For this purpose Wu and Huang suggested in 2009 a noise assisted data analysis method, called EEMD \citep{Wu2009}. For a given signal white noise is added to the signal and then decomposed using the EMD. At the end - after numerous trials - we take the mean of the IMF's. "By adding finite noise, the EEMD elimated largely the mode mixing problem and preserve physical uniqueness of decomposition. Therefore, the EEMD represents a major improvement of the EMD method." \citep{Wu2009}. Of course this is more time consuming than the usual EMD and we have to check if this is worthwhile in our application or not.
	
\subsection{Sparse Decomposition}
	The last method we would like to present is the Sparse Decomposition. We assume that the given data is a superposition of some processes. Then we can find these by minimizing the residual
	\begin{align*}
		\| \vec k_1 x_1 + \ldots + \vec k_n x_n - \vec y \|_2,	
	\end{align*}
	where $\vec k_1, \ldots, \vec k_n$ are the possible patterns. Writing $\vec k_1, \ldots, \vec k_n$ as the columns of a matrix $K$, then this is the same as 
	\begin{align*}
		\| K \vec x - \vec y \|_2.	
	\end{align*}
	The Fourier and the Wavelet Transform can also be written as such a minimization problem by choosing the patterns as harmonics or wavelets respectivly. The advantage compared to the Fourier and Wavelet Transform is that we are not restricted to only one basis to find our decomposition, but we can incorporate several bases, which is called a \textit{dictionary}. For our purposes we have chosen a dictionary with translational displacements and scalings of the pattern in figure \ref{RFSS:Dictionary}, hence we combine the capability of the Fourier and Wavelet Transform. Furthermore, we can also add any other pattern to increase the physical meaning of the decomposition.
 
	We are looking for a solution which has a small residual and is \textit{sparse}, i.e. it uses only a few patterns. A decomposition of the data into only a few patterns is likely to have physical meaning, especially when the patterns are chosen according to physical effects. This can be achieved by minimizing 
	\begin{align*}
		\tfrac{1}{2}\|K\vec x-\vec y\|_2^2 + \alpha \|\vec x\|_1, 
	\end{align*}
        see \cite{Daubechies2004}. The last term $\|\vec x\|_1$, called \textit{penalty term}, is the sum of the absolute values of $\vec x$. The notion of Sparse Decomposition is also investigated in an applied manner in \cite{Chen1999}.%

	For our application we have chosen the algorithm Regularized Feature Sign Search (RFSS), invented by \cite{Jin2009} and based on \cite{Lee2007}, to minimize these functional.

	\def\picwidth{0.19\textwidth}
	\def\date{2011-05-13}
	\begin{figure}[ht]
		\centering
		\subfigure[harmonics]
			{\includegraphics[width=\picwidth]{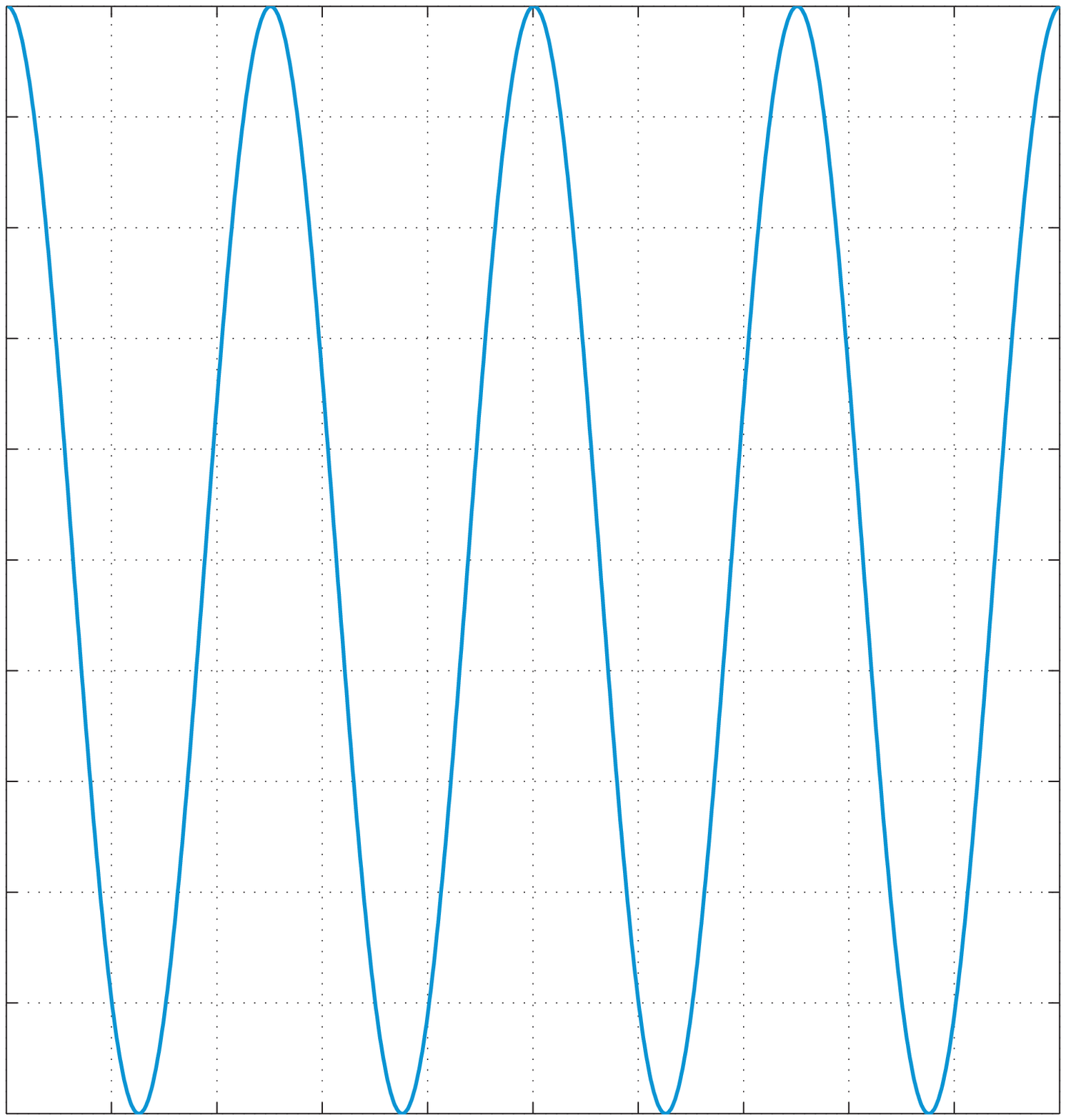}} 
		\subfigure[wavelets]
			{\includegraphics[width=\picwidth]{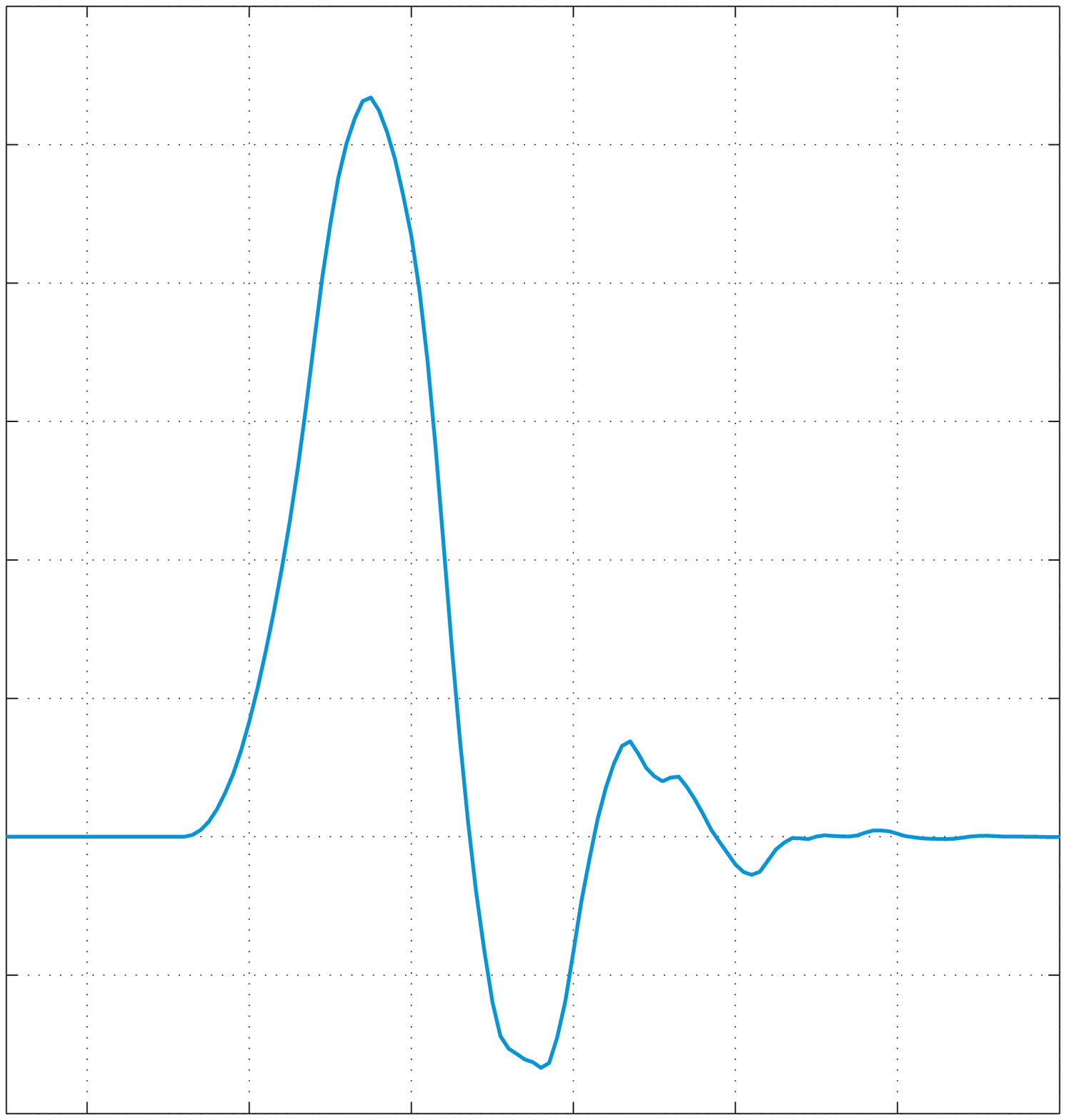}} 
		\subfigure[peaks]
			{\includegraphics[width=\picwidth]{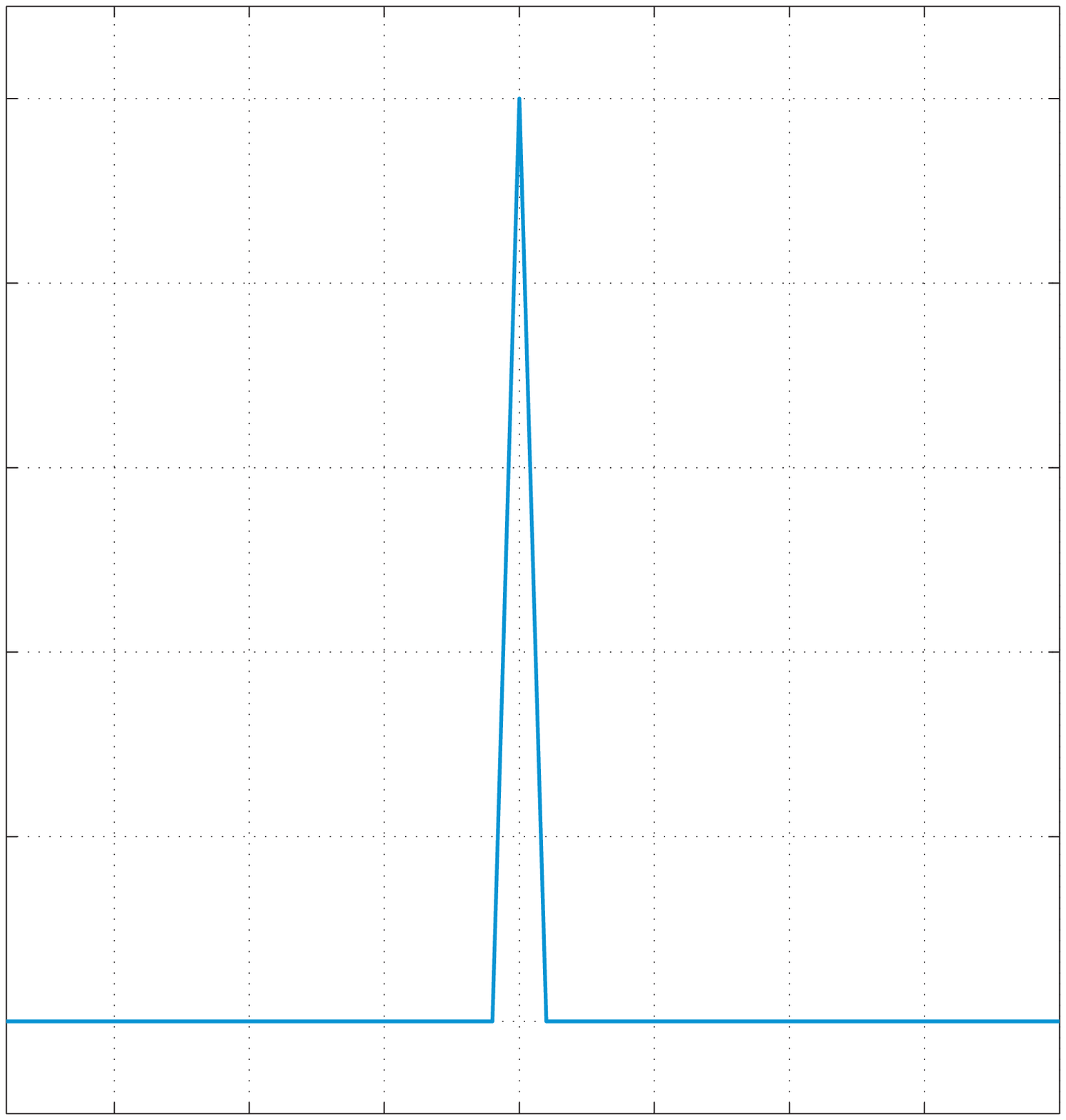}}
		\subfigure[blocks]
			{\includegraphics[width=\picwidth]{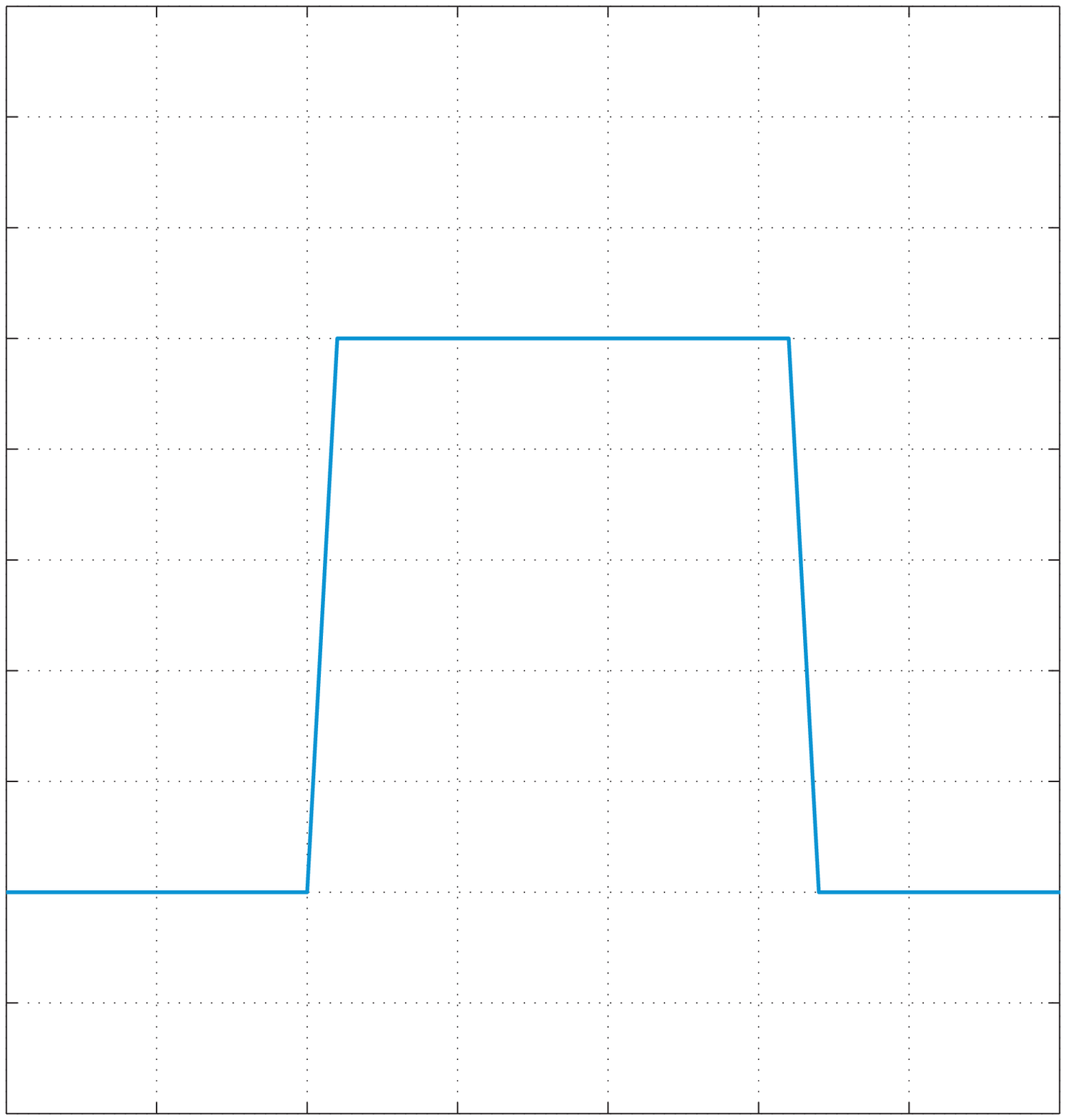}} 
		\subfigure[linear trend]
			{\includegraphics[width=\picwidth]{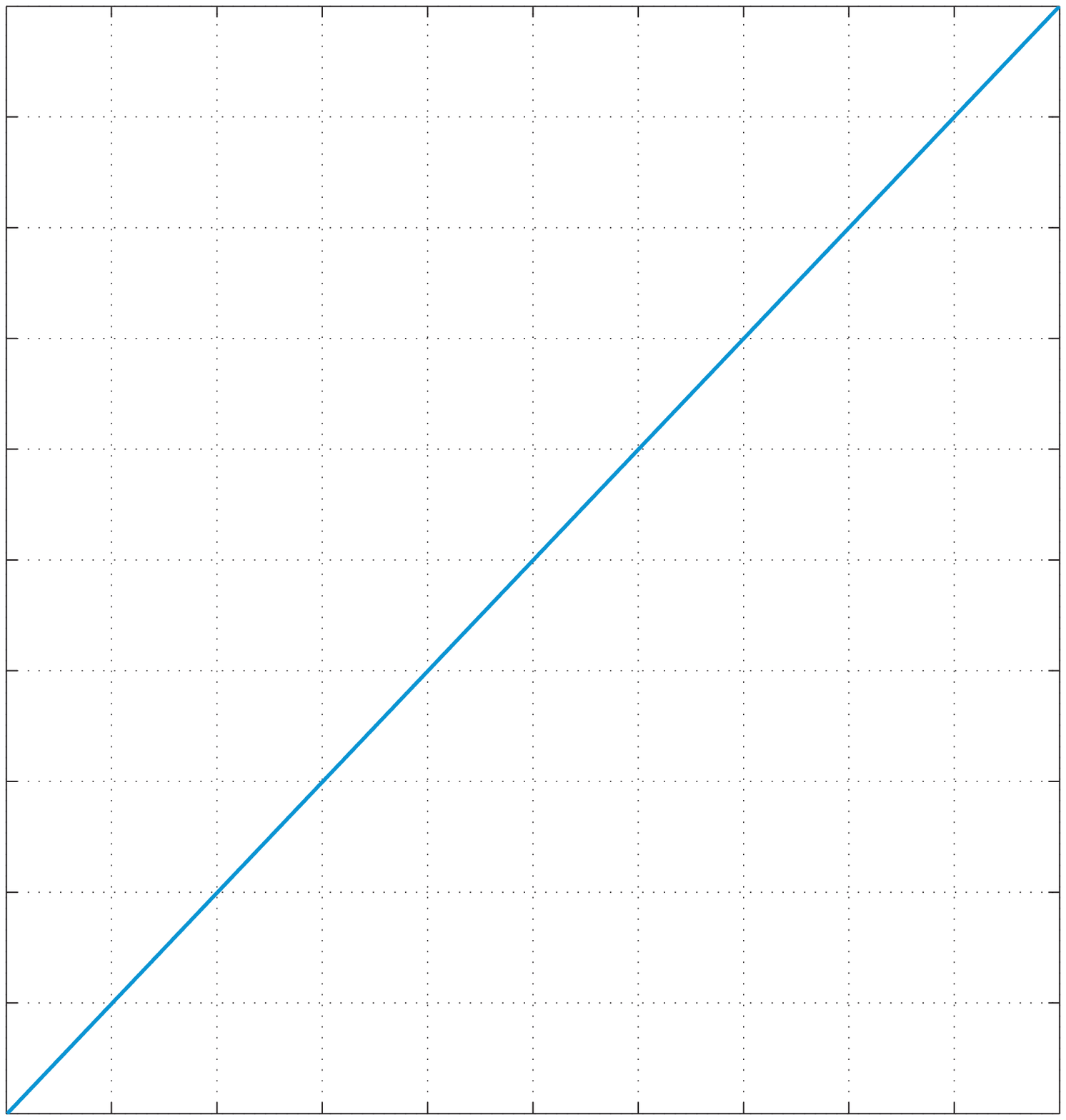}}
		\caption{The chosen dictionary which contains harmonic waves, a wavelet basis, a peak basis, blocks and a linear trend.}
		\label{RFSS:Dictionary}
	\end{figure}	

	The choice of the dictionary is an advantage and disadvantage at the same time. If we have special knowledge of our data, like in most cases, we can improve the decomposition by choosing 'good' pattern. On the other hand an algorithm which requires many input parameters is often difficult to handle. A disadvantage is that the algorithm is much more demanding and complex compared to the other methods. We will see later on that this is worthwhile. 

	Further information and the full algorithm is given by \cite{Jin2009}.

%% file: emd_chart.tex
\pgfdeclarelayer{background}
\pgfsetlayers{background,main}

\tikzstyle{decision} = [rectangle, 
			rounded corners=12pt,
			draw=black,
			fill=blue!0,
			text centered, 
			inner sep=5pt, 
			minimum width=2cm, 
			minimum height=0.8cm
]

\tikzstyle{block} = [	rectangle, 
			draw=black, 
			fill=blue!0, 
			text centered, 
			rounded corners=4pt, 
			minimum height=0.8cm,
			text width=2.9cm
]

\tikzstyle{line} = [draw, -latex']
    
\newlength{\HeightSmall}	\setlength{\HeightSmall}{3.6cm}
\def\HeightI{1.4cm}
\def\HeightII{2.4cm}
\def\HeightIn{1.5cm}
\def\HeightOut{1.8cm}
\def\WidthI{3.9cm}
\def\WidthII{3.5cm}

\begin{tikzpicture}
    \node [	block] (init) 
	{initialize};
    \node [	block, 
		below of=init, 
		node distance=\HeightSmall] (extrema) 
	{find extrema};
    \node [	decision, 
		below of=extrema, 
		node distance=\HeightI] (numberExtrema) 
	{\#extrema $\leq 1$}; %
    \node [	block, 
		below of=numberExtrema, 
		node distance=\HeightOut] (terminate) 
	{terminate};
    \node [	block, 
		right of=numberExtrema, 
		node distance=\WidthI] (envelope) 
	{compute mean value};
    \node [	decision, 
		above of=envelope, 
		node distance=\HeightI] (meanZero) 
	{mean $\approx 0$};
    \node [	block, 
		above of=meanZero, 
		node distance=\HeightI] (substractMean) 
	{substract mean};
    \node [	block, 
		right of=meanZero, 
		node distance=\WidthII] (FoundIMF) 
	{found IMF}; 
    \node [	block, 
		above of=FoundIMF, 
		node distance=\HeightII] (substractIMF) 
	{substract IMF};

    \path [line] (init) 		-- 					(extrema);
    \path [line] (extrema) 		-- 					(numberExtrema);
    \path [line] (numberExtrema) 	-- node [near start, left] {yes}  	(terminate);
    \path [line] (numberExtrema) 	-- node [near start, above, xshift=1pt] {no} 	(envelope);
    \path [line] (envelope) 		-- 					(meanZero);
    \path [line] (meanZero) 		-- node [near start, above, xshift=2pt] {yes}	(FoundIMF);
    \path [line] (meanZero) 		-- node [near start, left, yshift=1pt] {no}  	(substractMean);
    \path [line, rounded corners=10pt] (substractMean) 	-| 					(extrema);
    \path [line] (FoundIMF) 		-- 					(substractIMF);
    \path [line, rounded corners=10pt] (substractIMF) 	-| 					(extrema);

    \begin{pgfonlayer}{background}
        \path[fill=blue1!20, rounded corners, draw=black!50, dashed]
		(numberExtrema) + (-1.8cm, -1.0cm) rectangle (9.1cm, -0.7cm);
        \path[fill=blue1!40, rounded corners, draw=black!50, dashed]
            	(numberExtrema) + (-1.7cm, -0.9cm) rectangle (5.6cm, -1.7cm);
    \end{pgfonlayer}
\end{tikzpicture}

%% file: results.tex
\section{Results}

In this section we present and discuss our results. First of all we have to mention that the used data is preprocessed by subtracting the mean value, so the important features are emphasized.

Our results for Harmonic and Wavelet Decomposition are obtained by using Matlab\textsuperscript{\textregistered}'s functions \verb|fft| and \verb|wavedec| respectivly. For EMD and EEMD we have used the algorithms by \cite{Rilling2007} and \cite{Wu200X}. The RFSS implementation, used for the Sparse Decomposition, was according to \cite{Schiffler2009}. All algorithms can be found on the world wide web, see our references.

The tidal extraction from the data sets is done very well by all algorithms, hence this does not distinguish the methods and is not discussed in the following.

\subsection{Decomposition of SYN}
The results of the decomposition of the synthetic data are shown in figure \ref{RESULTS:SYN}. Since it is a synthetic data set we can compare the computed decomposition by the real one to evaluate their quality. 

As we see the methods' ability to decompose this data varies substantially. Harmonic and Wavelet Decomposition are able to decompose the data into  features of three different types. The noise and the short effects are in the same component. The methods EMD and EEMD are not able to extract the high frequency features from the signal. All algorithms were able to separate the long-term effects from the rest but failed to distinguish between the trend and the step. 

Harmonic and Wavelet Decomposition as well as the Sparse Decomposition correctly detect the short time features as well in time as in amplitude. Only large boundary effects in the results obtained by Harmonic and Wavelet degrade their quality. Methodically we can not obtain boundary effects in the results of the Sparse Decomposition, since we have chosen our dictionary to contain local and global pattern. Hence local features do not have to be reconstructed by global pattern. Additionally the RFSS is not based on filtering, which often causes boundary effects. Overall the Sparse Decomposition performs best in detecting the short time features.

The last aim was to extract the long time effects. All algorithms perform quite well, but again the boundary effects degrade the Harmonic and Wavelet results. Sparse Decomposition is capable of extracting the trend and the step very well whereas EMD and EEMD reproduce only the trend satisfactorily.

\def\picwidth{0.44\textwidth}

\begin{figure}[ht]	
	\centering
	$ $\\[-7mm]	
	\subfigure[Harmonic Decomposition]{
		\includegraphics[width = \picwidth]{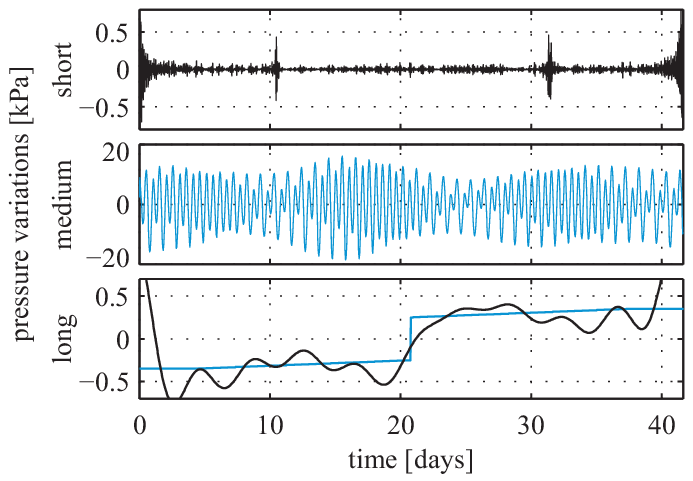}} \hfill
	\subfigure[Wavelet Decomposition]{
		\includegraphics[width = \picwidth]{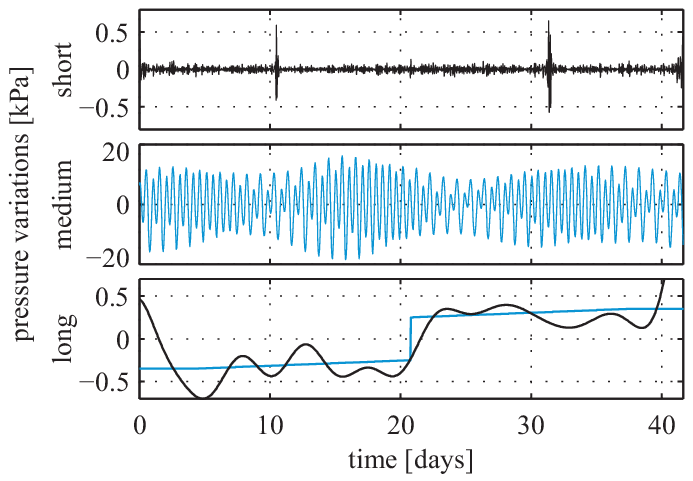}} \\
	\subfigure[EMD]{
		\includegraphics[width = \picwidth]{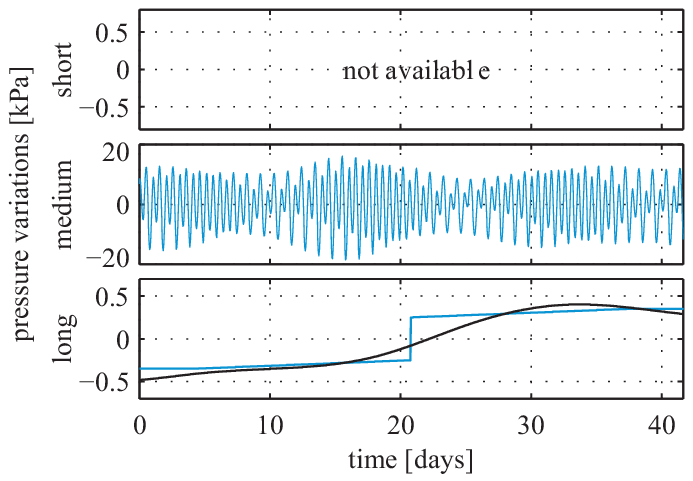}} \hfill
	\subfigure[EEMD]{
		\includegraphics[width = \picwidth]{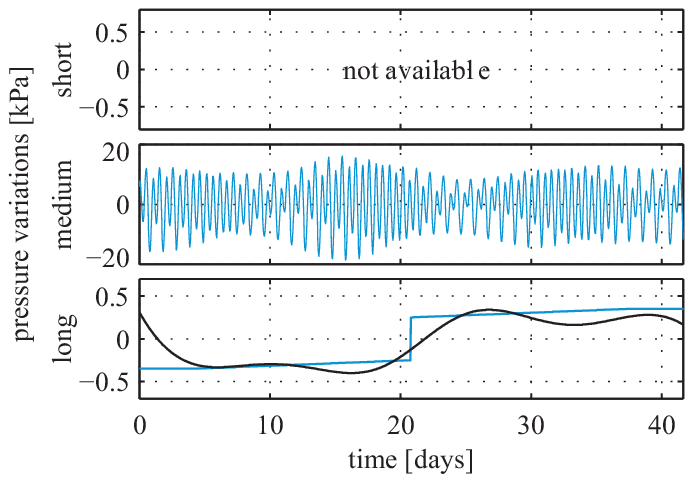}} \\
	\subfigure[Sparse Decomposition]{
		\includegraphics[width = \picwidth]{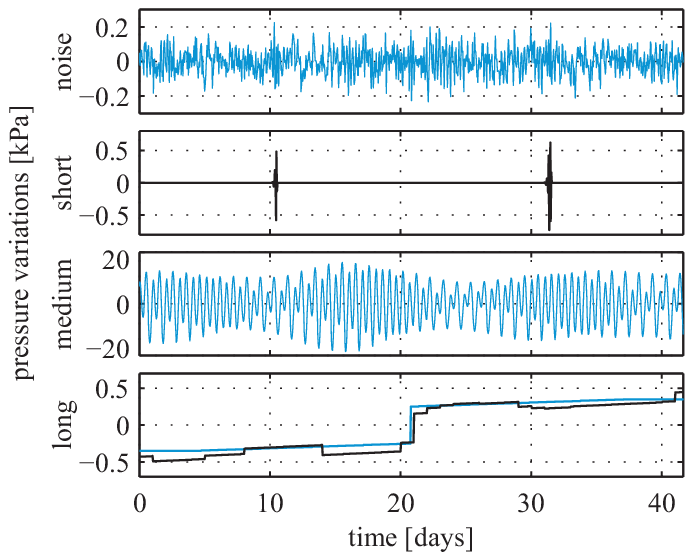}} \hfill     
	\def\name{Synthetic_short_data2}
	\def\date{2011-07-22}
	\subfigure[synthetic data]{
		\includegraphics[width = \picwidth]{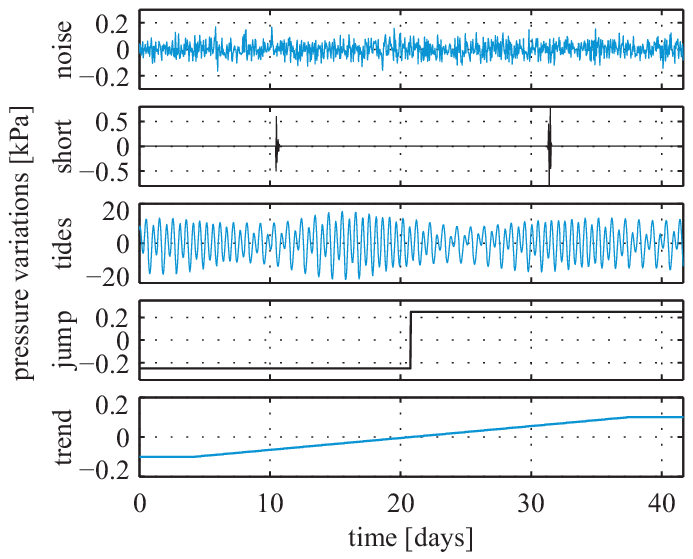}}
	\caption{Decomposition of data set SYN. There are shown from the top to the bottom the short, medium and long time components. The real decomposition can be seen in figure \ref{RESULTS:SYN}(f). EMD and EEMD were not capable to detect features with a higher frequency than the tides.}
	\label{RESULTS:SYN}
\end{figure}

\subsection{Decomposition of MAR}
\begin{figure}[ht]	
	\centering
	$ $\\[-7mm]
	\subfigure[Harmonic Decomposition]{
		\includegraphics[width = \picwidth]{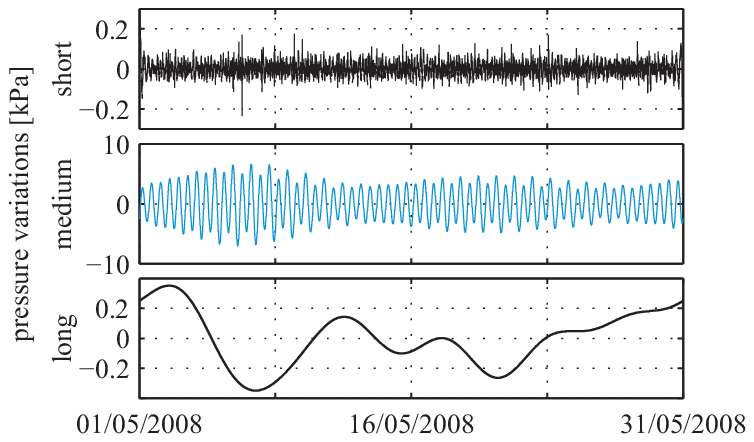}} \hfill
	\subfigure[Wavelet Decomposition]{
		\includegraphics[width = \picwidth]{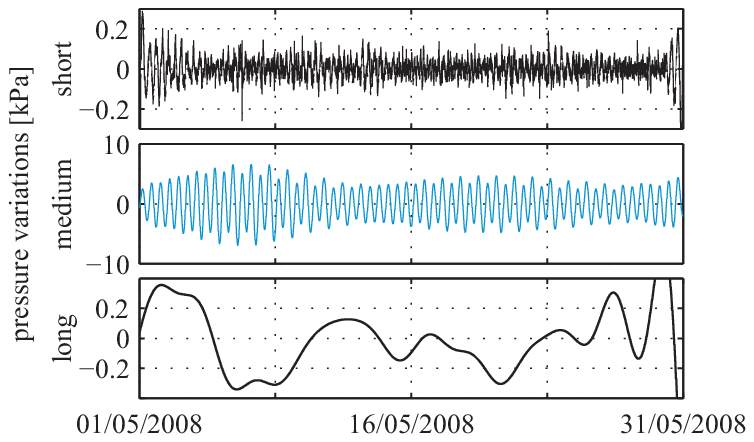}} \\
	\subfigure[EMD]{
		\includegraphics[width = \picwidth]{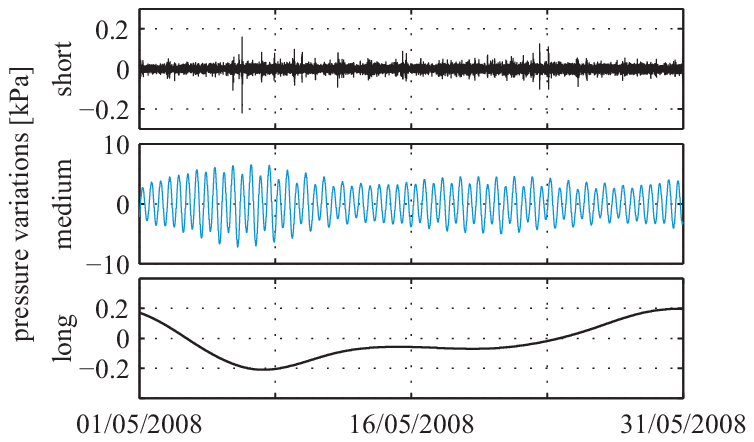}} \hfill
	\subfigure[EEMD]{
		\includegraphics[width = \picwidth]{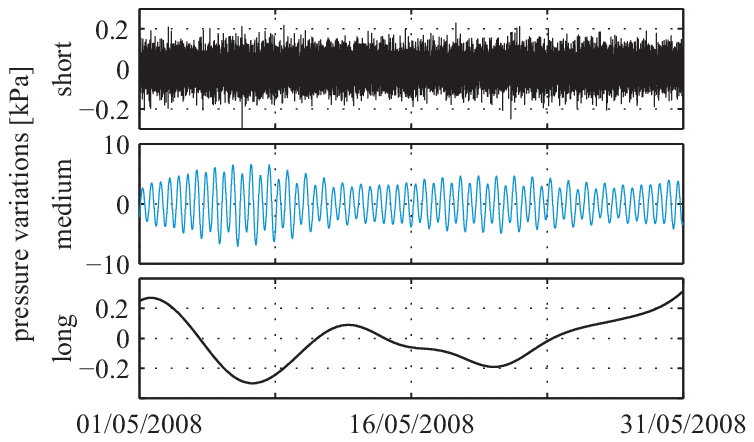}} \\
	\flushleft
	\subfigure[Sparse Decomposition]{
		\includegraphics[width = \picwidth]{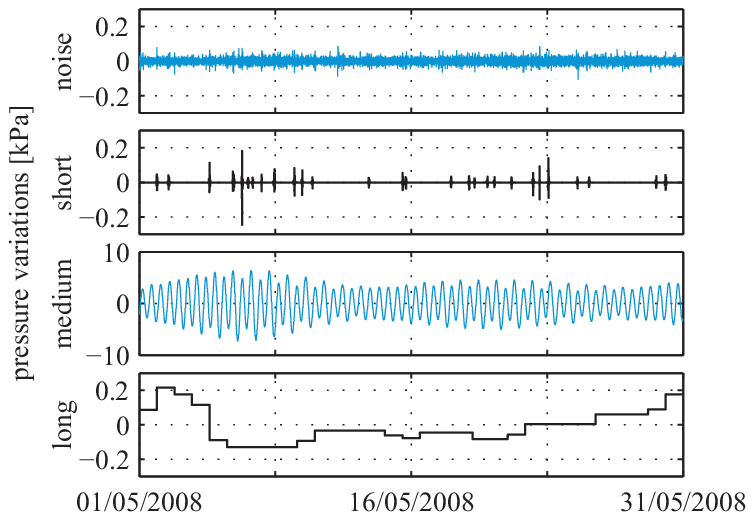}} 
	\caption{Decomposition of the data set MAR. There are shown from the top to the bottom the short, medium and long time components.}	
	\label{RESULTS:MAR}
\end{figure}

The next data set to test is the one from the Mid-Atlantic Ridge and the results can be seen in figure \ref{RESULTS:MAR}. This time all algorithms provide a good decomposition into the short-, medium- and long-period features. Only the Sparse Decomposition is able to extract also the noise from the short-period components. There are just a few peaks visible in the short-period decomposition by Harmonic Decomposition and EMD which are probably aliased local eartquake signals. These are also extracted very well by the Sparse Decomposition. The short-period features given by Wavelet show again some boundary effects. The decomposition by the EEMD has a basic noise level which is much too high. This is a result of the white noise added in every step of the algorithm but which is clearly not faded away with 100 runs of the EMD. 

The trend and long-term feature extraction of the five methods are all nearly the same. In all cases pressures decrease initially over the first week and then increase by about 0.4 $kPa$ over the rest of the time window. The Wavelet results differ from the others by significant boundary effects.

\subsection{Decomposition of CORK1}%

\begin{figure}[ht]	
	\centering
	$ $\\[-7mm]
	\subfigure[Harmonic Decomposition]{
		\includegraphics[width = \picwidth]{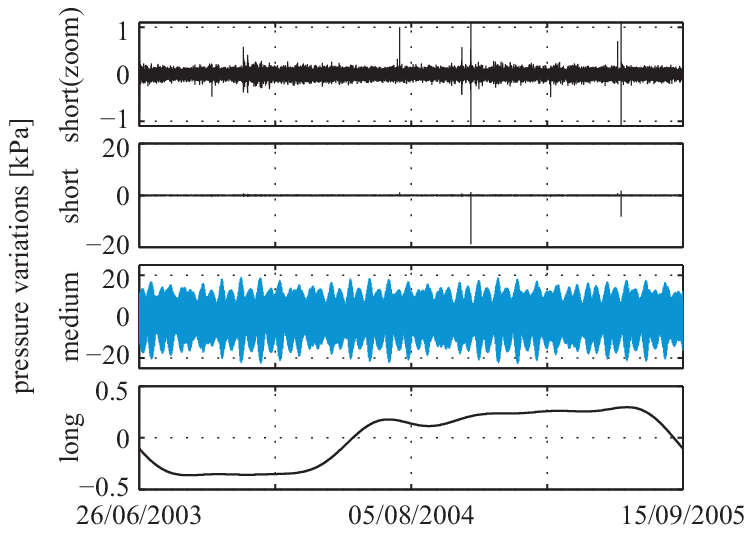}} \hfill
	\subfigure[Wavelet Decomposition]{
		\includegraphics[width = \picwidth]{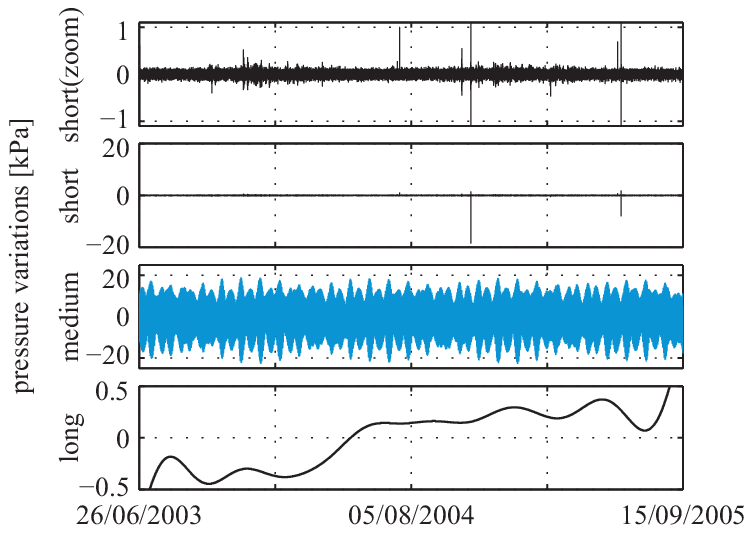}} \\
	\subfigure[EMD]{
		\includegraphics[width = \picwidth]{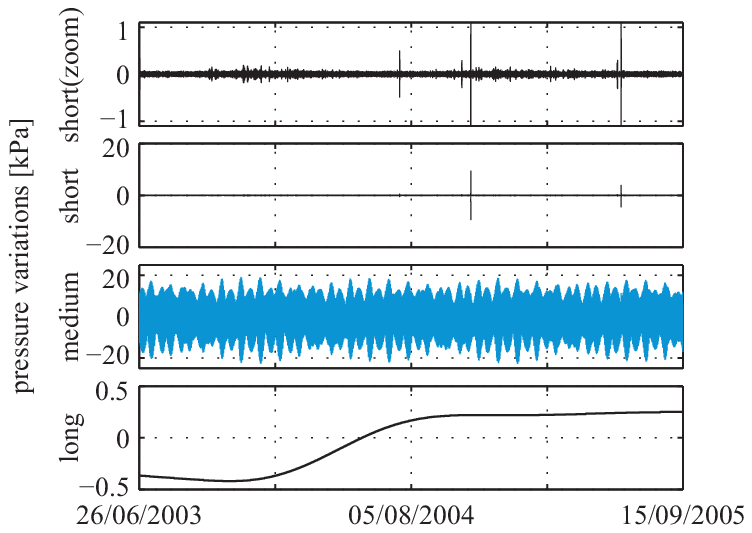}} \hfill
	\subfigure[EEMD]{
		\includegraphics[width = \picwidth]{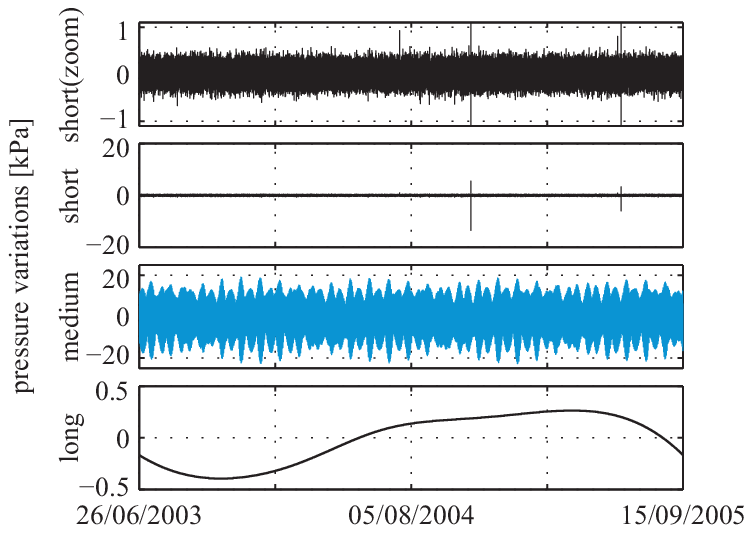}} \\
	\flushleft
	\subfigure[Sparse Decomposition]{
		\includegraphics[width = \picwidth]{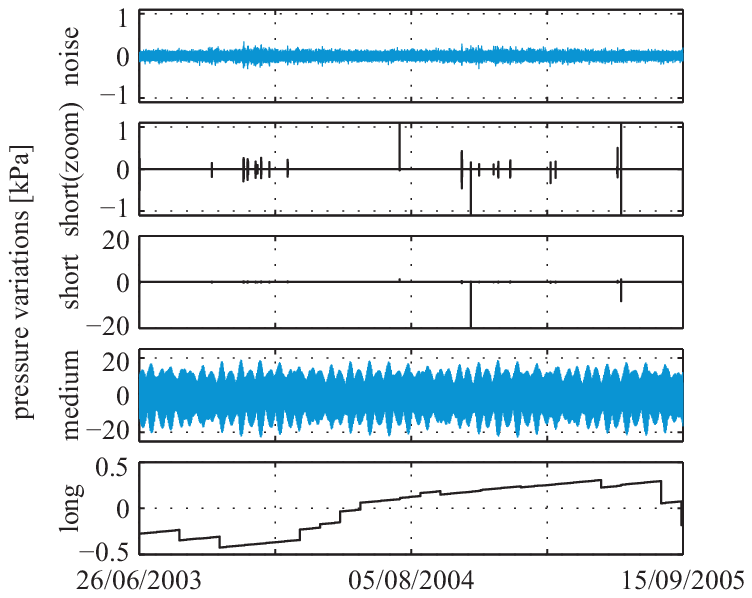}}	
	\caption{Decomposition of the data set CORK1. There are shown from the top to the bottom the short, medium and long time components.}
	\label{RESULTS:CORK1}	
\end{figure}

The results of the third data set, CORK1, are given in figure \ref{RESULTS:CORK1}. This time every method is again capable to decompose the data into the three main features. In this data set there are two short time effects whose amplitudes are a lot higher than the amplitudes of the other effects so we have also showed the short period effects zoomed in.%

The similarity between the short period features detected by Harmonic and Wavelet Decomposition is worth mentioning. The EMD and EEMD detect these features very similar in time but differ in detecting the amplitude. Also the EEMD is not able to detect short period features with a small amplitude since the added noise prevents this. The Sparse Decomposition is able to detect these features as well and separates them additionally from the noise.%

The trends detected by all methods are again quite similar. Only the Wavelet trend has huge boundary effects. The behaviour of Harmonic Decomposition at the boundary could be interpreted as a boundary effect, but we guess this is not the case, since Sparse Decomposition predict the same and is methodically unaffected by boundary effects. The same behaviour can be obtained from the EEMD's trend.%

\subsection{Decomposition of CORK2}%
\begin{figure}[ht]%
	\centering
	$ $\\[-7mm]
	\subfigure[Harmonic Decomposition]
		{\includegraphics[width = \picwidth]{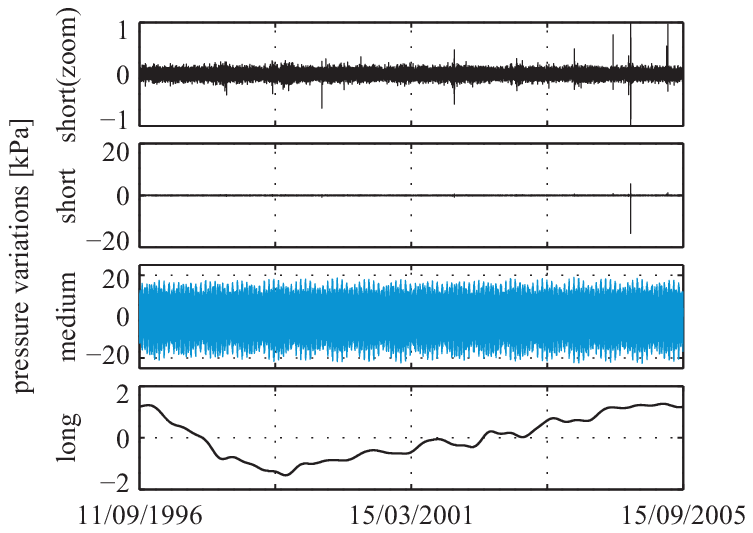}} \hfill
	\subfigure[Wavelet Decomposition]
		{\includegraphics[width = \picwidth]{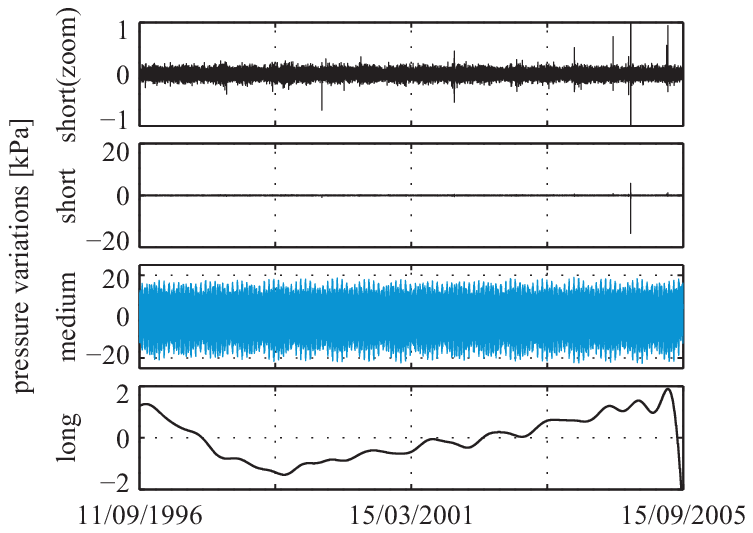}} \\
	\subfigure[EMD]
		{\includegraphics[width = \picwidth]{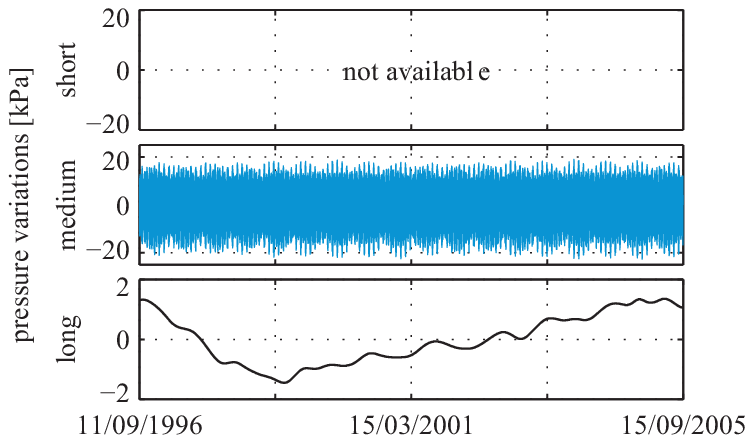}} \hfill
	\subfigure[EEMD]
		{\includegraphics[width = \picwidth]{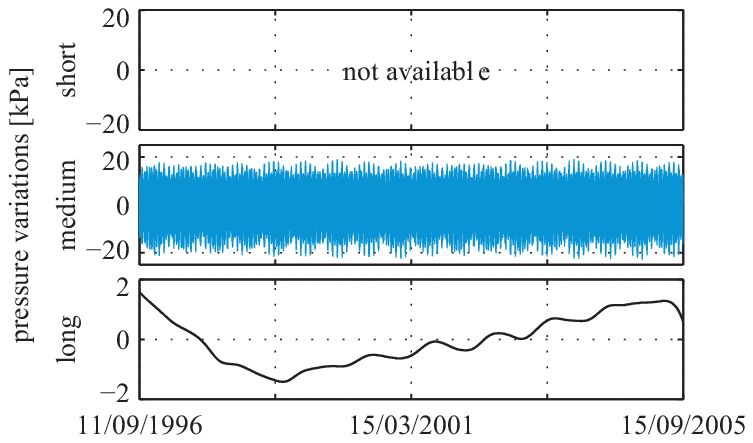}} \\
	\flushleft
	\subfigure[Sparse Decomposition]
		{\includegraphics[width = \picwidth]{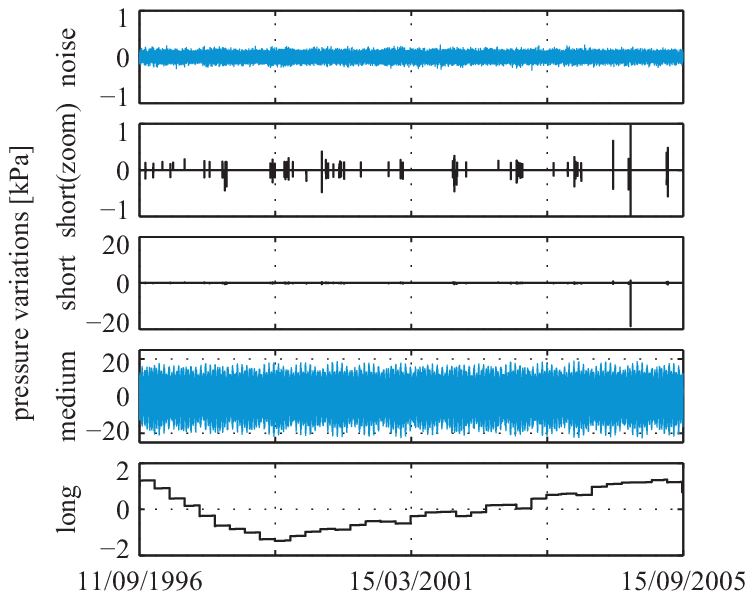}}
	\caption{Decomposition of the data set CORK2. There are shown from the top to the bottom the short, medium and long time components. Since the short time components has features of different amplitudes these are also showed zoomed in. EMD and EEMD failed to detect the short period features.}
	\label{RESULTS:CORK2}	
\end{figure}

For the last data set the results are given in figure \ref{RESULTS:CORK2}. Since this is a long data set and the sampling interval is only 60 $min$ we focus on the long period effects. On the other hand we can not expect short time features with a shorter duration than 60 $min$, like earthquakes, to appear in this data set, hence the short time components obtained from this data set has to be interpreted very carefully.%

Like with the dataset SYN, which sampling interval was also 60 $min$, EMD and EEMD are not capable to decompose the features with higher frequencies than the tides. Contrary this is done very well by Harmonic, Wavelet and Sparse Decomposition. There are one huge and some smaller short period signals associated to earthquakes, which are again detracted from the noise by Sparse Decomposition.%

The trends detected by all methods look quite similar. In any case we obtain first a downdrift of 3.5 $kPa$ over 2 years, followed by an updrift the same amount in the next seven years. As seen at the other data sets the boundary effects are very high at the Wavelet Decomposition.%

\subsection{Discussion}

A summary of the observations described in the previous sections is given in table \ref{tab:results}. The evaluation of the different decomposition methods was based on the following more or less qualitative criteria:
\begin{itemize}
	\item How successful was the algorithm in decomposing the signal into a long and short period components?
	\item Did the algorithm find a short period event in the presence of background noise?
	\item Does the algorithm create large effects at the boundaries of the time window?
	\item How much CPU time is needed for the decomposition?
\end{itemize}
\def\tinySpace{1mm}
\def\smallSpace{3mm}
\def\largeSpace{5mm}
\def\width{22mm}
\def\widthDouble{44mm}
\begin{table}[ht]
	\centering
	\caption[Overview of our results]{Overview of our results. \good \;indicates good, \medium \;fair, and \bad \;bad results.}
	\begin{tabular}{cccccccc}
		\toprule
		& &		& Harmonic	& Wavelet	& EMD		& EEMD	 	& Sparse	\\ 
		\midrule
	\multirow{8}{*}{\rotatebox{90}{\parbox{\widthDouble}{\centering Decomposition}}} 
		& \multirow{4}{*}{\rotatebox{90}{\parbox{\width}{\centering short period}}}  
		  & SYN	   	& \good  	& \good 	& \bad		& \bad   	& \good		\\[\tinySpace] 
                & & MAR	   	& \medium   	& \medium 	& \good		& \bad  	& \good 	\\[\tinySpace]  
                & & CORK1	& \good		& \good	   	& \medium	& \medium	& \good		\\[\tinySpace]  
 		& & CORK2	& \good 	& \good		& \bad		& \bad		& \good		\\[\smallSpace]     
        	& \multirow{4}{*}{\rotatebox{90}{\parbox{\width}{\centering long period}}}   
		  & SYN		& \medium  	& \medium	& \good		& \medium	& \good		\\[\tinySpace]  
                & & MAR		& \good 	& \medium	& \good		& \good 	& \good		\\[\tinySpace] 
                & & CORK1	& \good		& \medium	& \good		& \good		& \good		\\[\tinySpace]  
 		& & CORK2	& \medium  	& \medium   	& \good		& \good		& \good		\\[\largeSpace]   
       \multirow{4}{*}{\rotatebox{90}{\parbox{\width}{\centering Boundary}}}
		&\multirow{4}{*}{\rotatebox{90}{\parbox{\width}{\centering effects}}} 
		  & SYN		& \bad  	& \bad 		& \good		& \medium	& \good		\\[\tinySpace]  
                & & MAR		& \medium 	& \bad	   	& \good		& \good		& \good		\\[\tinySpace]  
                & & CORK1	& \medium 	& \bad	   	& \good		& \good		& \good		\\[\tinySpace] 
 		& & CORK2	& \medium	& \bad		& \good		& \good		& \good		\\[\largeSpace]          \multicolumn{3}{c}{Computing efficiency}  
                		& \good 	& \good 	& \medium 	& \bad 		& \bad		\\ 
	\bottomrule
	\end{tabular} 
	\label{tab:results}
\end{table}

From looking at table \ref{tab:results} one can conclude that none of the used methods for decomposition is perfect in every aspect.
Harmonic and Wavelet are able to detect the short period features and perform also quite well in detracting the trend but their results, especially the results of Wavelet, show massive boundary effects. Because of this there is a high uncertainty if the shown decomposition is physically reasonable or not. On the other hand their computing efficiency is so high that the Harmonic and Wavelet Decomposition is nearly not restricted by the length of the given data.

The performance of EMD and EEMD is directly opposite to Harmonic and Wavelet. The EMD and EEMD perform very well on detracting the tides and show only small influences of boundary effects but their capability of separating components with higher frequencies than the tides is restricted. At least when the data is recorded with a sampling interval of 60 minutes. Especially the EEMD fails because the added noise does not fade away in our experiments. Overall we have seen in every data set that the large additional computational effort of the EEMD compared to the EMD is not worthwhile for the analysis of sea floor pressure data.

Very impressive is the capability of the Sparse Decomposition. It was able to detract the short period components as well as the large period components in every case. Contrary to the other methods, it was also able to detract the noise from the meaningful short period components. A hugh advantage is also that the Sparse Decomposition does not have any boundary effects. Only the computing efficiency needs to be improved.

\clearpage

%% file: conclusion.tex
\section{Conclusion}

To conclude our investigation we have to state that no method is perfect for every time series. For our data sets we got the best results by using the Sparse Decomposition. But at the moment it is not recommendable for time series with more than 200,000 measurements. For this kind of time series we recommend the EMD for long period contributions and the Harmonic Decomposition for the short ones. EEMD fails to improve the EMD in detecting and is not recommended due to ist large computational needs. We obtained sometimes good results using the Wavelet Decomposition but they were never better than the results obtained by the Harmonic Decomposition, even in detecting earthquakes which is classically done by Wavelets. 

It is obvious that the comparison of the computing efficiency is somewhat unfair as algorithms to calculate the FFT and Wavelet Transform are highly optimized whereas algorithms for EMD, EEMD and especially RFSS are not yet written for speed. However it is very likely that FFT and Wavelet Transform will always be faster however the difference to EMD, EEMD and RFSS will most likely shrink in the near future due to the improvement of the algorithms and maybe by developing parallel algorithms for them.

%% file: acknowledgements.tex
\section{Acknowledgements}

We would like to thank Hans-Hermann Gennerich and Earl Davis for providing the sea floor pressure data and Peter Maass for supporting one of us (ME) during this study. All computations were done on the computer network provided by the Department of Mathematics and Computer Science of the University of Bremen. Their help and support is kindly acknowledged. Reviews of two anonymous reviewers helped to improve the article.

%% file: references.tex
\bibliographystyle{model2-names}

\bibliography{references}
\addcontentsline{toc}{section}{References}
\nocite{*}